\documentclass[sigconf]{acmart}

\usepackage{xcolor}
\usepackage{verbatim}
\usepackage{graphicx}
\usepackage{tikz}
\usetikzlibrary{arrows.meta, positioning, shapes.geometric}

\AtBeginDocument{%
  \providecommand\BibTeX{{%
    \normalfont B\kern-0.5em{\scshape i\kern-0.25em b}\kern-0.8em\TeX}}}

\setcopyright{acmlicensed}
\copyrightyear{2024}
\acmYear{2024}
\acmDOI{XXXXXXX.XXXXXXX}

\title{Towards a Characterization of Microservice Architectures Generated by Large Language Models}

\author{José Renan Alves Pereira}
\email{jose.pereira@embedded.ufcg.edu.br}
\affiliation{%
  \institution{VIRTUS/UFCG, Federal University of Campina Grande}
  \city{Campina Grande}
  \state{Paraíba}
  \country{Brazil}
}

\author{Ademar Sousa}
\email{ademar.sousa@virtus.ufcg.edu.br}
\affiliation{%
  \institution{VIRTUS/UFCG, Federal University of Campina Grande}
  \city{Campina Grande}
  \state{Paraíba}
  \country{Brazil}
}

\author{Emanuel Dantas Filho}
\email{emanuel.dantas@virtus.ufcg.edu.com.br}
\affiliation{%
  \institution{Federal University of Campina Grande - ISE/VIRTUS}
  \city{Campina Grande}
  \state{Paraíba}
  \country{Brazil}
}

\author{Danyllo Albuquerque}
\email{danyllo.albuquerque@virtus.ufcg.edu.br}
\affiliation{%
  \institution{VIRTUS/UFCG, Federal University of Campina Grande}
  \city{Campina Grande}
  \state{Paraíba}
  \country{Brazil}
}

\author{Mirko Perkusich}
\email{mirko@virtus.ufcg.edu.br}
\affiliation{%
  \institution{VIRTUS/UFCG, Federal University of Campina Grande}
  \city{Campina Grande}
  \state{Paraíba}
  \country{Brazil}
}

\author{Kyller Gorgônio}
\email{kyller@virtus.ufcg.edu.br}
\affiliation{%
  \institution{VIRTUS/UFCG, Federal University of Campina Grande}
  \city{Campina Grande}
  \state{Paraíba}
  \country{Brazil}
}

\author{Angelo Perkusich}
\email{perkusic@virtus.ufcg.edu.br}
\affiliation{%
  \institution{VIRTUS/UFCG, Federal University of Campina Grande}
  \city{Campina Grande}
  \state{Paraíba}
  \country{Brazil}
}

\setcopyright{none}
\acmYear{2026}
\acmDOI{}
\acmConference[JAWs '26]{Journal Ahead Workshop}{2026}{}
\acmBooktitle{Journal Ahead Workshop (JAWs 2026)}

\begin{document}

\begin{abstract}
\textbf{Background:} Large Language Models (LLMs) are increasingly used for software design tasks, yet little is known about the structure and coherence of the architectures they produce. In particular, how LLMs behave in Architecture-to-Architecture transformations remains insufficiently understood. \textbf{Objective:} This study empirically characterizes how LLMs generate microservice architectures from natural language descriptions of modular monoliths, focusing on structural and descriptive properties rather than architectural correctness. 
\textbf{Method:} We conduct controlled experiments on two modular monolith systems, comparing zero-shot and few-shot prompting strategies using GPT-based models from OpenAI and models from DeepSeek. Generated architectures are represented in a standardized CSV format and evaluated using normalized metrics capturing service granularity, inter-service communication patterns, communication density, service isolation, and responsibility descriptiveness. \textbf{Results:} The results reveal systematic differences driven primarily by prompting strategy rather than model provider. Few-shot prompting consistently produces finer-grained decompositions with lower communication density and more detailed responsibility descriptions, whereas zero-shot prompting favors coarser and more tightly connected architectures. \textbf{Conclusion:} Rather than assessing architectural quality, this study provides an empirical baseline for understanding LLM behavior in text-driven Architecture-to-Architecture transformations, informing future research on generative design workflows and hybrid architectural reasoning in early-stage software modernization.
\end{abstract}

\maketitle





\section{Introduction}

The widespread adoption of cloud computing has intensified the demand for software systems that can evolve rapidly and respond efficiently to changes in infrastructure and business requirements. In this context, microservices have emerged as a prominent architectural style, enabling scalability, flexibility, and independent deployment of system components~\cite{slr_mono_micro, defining_micro_and_mono, 62_neural_network, mono_to_micro_challenges, 65}. Despite their advantages, many software systems continue to rely on monolithic architectures, which, although historically successful, are increasingly constrained by centralized deployment, tight coupling, and limited adaptability~\cite{18, 68, 1807.10059v2}.

Consequently, the decomposition of monolithic systems into microservices has become a central challenge in software architecture research and practice. Existing approaches predominantly rely on static analysis, dynamic analysis, clustering techniques, or combinations thereof, often requiring access to source code and execution traces~\cite{63, 60, 72}. While effective in specific contexts, these approaches may be impractical in early modernization scenarios, where detailed implementation artifacts are unavailable or incomplete. In such early-stage settings, architectural reasoning often relies on documentation artifacts (e.g., SADs) and modular views rather than complete implementation data.

More recently, Large Language Models (LLMs) have attracted growing attention in software engineering due to their ability to process and reason over textual artifacts~\cite{WL30, ESPOSITO2026112607}. Prior studies have explored the use of LLMs for tasks such as requirements analysis, architectural decision support, and code generation. However, the application of LLMs to Architecture-to-Architecture transformations—specifically, the decomposition of monolithic systems into microservices based on textual descriptions—remains underexplored and lacks methodological consolidation~\cite{WL33, WL36, req_to_archi, ESPOSITO2026112607}.

In particular, existing studies in this area are still limited in number and scope, often employing heterogeneous techniques and evaluation strategies that hinder systematic comparison and replication~\cite{ESPOSITO2026112607, WL30, WL26}. Moreover, some approaches that rely on textual descriptions continue to depend on traditional machine learning or clustering methods, rather than thoroughly investigating the generative and reasoning capabilities of LLMs~\cite{63, 60, 72}. This fragmentation reveals a clear gap regarding reproducible methodologies and empirical analyses for text-driven monolith-to-microservice decomposition.

Despite growing interest in GenAI for software architecture, there is still limited empirical evidence characterizing how LLMs behave when performing Architecture-to-Architecture transformations based solely on textual descriptions and artifact-based evaluation.

To address this gap, this study investigates LLM-based microservice derivation from textual descriptions of modular monoliths by analyzing the generated specifications through an artifact-based, normalized approach.
 Specifically, we analyze two monolithic systems, MediaStore (MS) and TeaStore (TS), studied initially by Fuchs et al.~\cite{WL11}, which provide structured Software Architecture Documentation (SAD) artifacts. Using these textual descriptions as input, we employ zero-shot and few-shot prompting strategies with models from OpenAI and DeepSeek to generate high-level microservice architectures intended to support early-stage architectural modernization.

This work builds upon and extends a prior study by the authors that explored microservice recommendation from textual descriptions~\cite{sbcars}. In contrast to previous work, the present study considers multiple systems and multiple LLM providers, and a refined set of quantitative evaluation metrics designed to characterize the structural and descriptive properties of the generated architectures. The experimental setup reuses and adapts existing prompt templates and tooling, enabling systematic comparison across configurations.

To structure the investigation, we adapt the Monolith-to-Microservices Decomposition Framework (M2MDF) proposed by Abgaz et al.~\cite{slr_mono_micro}, modifying its workflow to accommodate LLM-based inference in place of traditional clustering or rule-based techniques. This adaptation provides a transparent, reusable methodological foundation for studying text-driven architecture decomposition with LLMs, contributing to greater methodological consistency in an area that currently lacks standardized practices.

Beyond its immediate empirical contribution, this study is intentionally positioned as an initial and foundational step within a broader research agenda on LLM-supported \emph{Architecture-to-Architecture} transformation for early-stage modernization. Rather than aiming to establish architectural correctness or prescribe a single ``best'' decomposition, the study focuses on characterizing systematic behavioral tendencies observed in LLM-generated microservice specifications when models are driven exclusively by textual architectural descriptions. The results clarify how prompting strategies and model configurations shape key architectural signals---such as service granularity, inter-service communication patterns, and responsibility descriptiveness---and expose trade-offs that are likely to recur in documentation-driven modernization contexts. This empirical grounding provides a necessary baseline for subsequent research stages, including the study of generation stability under repeated runs, the incorporation of additional evidence sources (e.g., static dependencies or runtime traces) to constrain or validate generated structures, and the alignment of artifact-based signals with expert architectural judgment through controlled human-in-the-loop studies. Together, these steps are intended to progressively converge toward more robust, explainable, and reproducible generative workflows for architecture modernization.

The remainder of this paper is organized as follows. Section~\ref{sec:background} reviews the foundations of monolith decomposition and situates recent advances in Generative AI for software architecture. Section~\ref{sec:methodology} details the research methodology, including the dataset, the adaptation of the M2MDF workflow, the prompting strategies, and the evaluation metrics. Section~\ref{sec:results} reports and discusses the experimental results. Section~\ref{sec:implications} discusses implications for researchers and practitioners, while Section~\ref{sec:threats} outlines threats to validity. Finally, Section~\ref{sec:finalremarks} concludes the paper and highlights directions for future work.

\section{Background}
\label{sec:background}
The migration from monolithic architectures to microservices is widely studied as a strategy to improve scalability, maintainability, and deployment agility. However, identifying appropriate service boundaries remains a persistent challenge, particularly in legacy systems where architectural knowledge is fragmented, outdated, or poorly documented~\cite{slr_mono_micro, 1807.10059v2, mono_to_micro_challenges}. In such settings, suboptimal decomposition decisions may increase coupling, accelerate architectural erosion, and compromise long-term maintainability.

Existing research typically classifies monolith decomposition techniques into static, dynamic, and semantic approaches~\cite{slr_mono_micro}. Static and dynamic methods rely on source code analysis and runtime traces to infer structural and behavioral dependencies~\cite{62, 66, 67, 68}, while semantic approaches apply Natural Language Processing techniques to align software components with business concepts and responsibilities~\cite{63, 72}. Although effective in implementation-rich scenarios, these techniques commonly assume access to source code, execution environments, or expert input, which is often unavailable during early-stage modernization efforts.

More recently, advances in Generative AI and Large Language Models (LLMs) have expanded the range of software architecture tasks that can be supported through automated reasoning over textual artifacts~\cite{ESPOSITO2026112607, WL26, WL6}. Prior studies report promising results in activities such as requirements interpretation, architectural decision support, and reverse engineering, often leveraging techniques such as Few-Shot Prompting, Chain-of-Thought reasoning, and Retrieval-Augmented Generation~\cite{ESPOSITO2026112607, WL26, WL6, WL18}. Despite this progress, the use of LLMs for \emph{Architecture-to-Architecture} transformations—particularly monolith-to-microservice decomposition driven by textual descriptions—remains comparatively underexplored and methodologically fragmented~\cite{ESPOSITO2026112607, WL30}.

\subsection{Monolith Decomposition and Generative AI}

A systematic review by Abgaz et al.~\cite{slr_mono_micro} highlights fragmentation and a lack of standardization in research on monolith-to-microservices decomposition. The authors propose the M2MDF, which structures existing approaches into six phases: input collection, monolith analysis, identification, optimization, evaluation, and deployment. Their findings indicate that most studies rely predominantly on static analysis techniques, focus mainly on Java-based systems, and assess results using coupling and cohesion metrics. The review also identifies critical gaps, including limited integration of heterogeneous data sources, scarcity of public datasets, and insufficient support for reproducibility.

Complementing this perspective, a Multivocal Literature Review by Esposito et al.~\cite{ESPOSITO2026112607} examines the emerging use of GenAI in software architecture. It reveals that the field is still in a nascent stage. The study reports an intense concentration on the GPT model family, widespread adoption of Few-Shot Prompting, and a predominant human-in-the-loop paradigm. Most existing work focuses on early pipeline stages, such as Requirements-to-Architecture and Architecture-to-Code transitions, whereas Architecture-to-Architecture scenarios—such as migration and refactoring—account for only a small fraction of the literature. Moreover, the review exposes severe methodological limitations, including the absence of formal validation methods (e.g., ATAM), the use of superficial evaluation criteria, and insufficient documentation of prompt design and experimental settings, collectively undermining reproducibility.

Another underexplored aspect concerns the use of textual architectural descriptions as primary inputs for decomposition tasks. In practice, software modernization initiatives frequently begin with partial documentation, architectural overviews, or requirement narratives rather than complete source code or runtime artifacts~\cite{defining_micro_and_mono}. Despite this reality, few studies systematically investigate how LLMs interpret and transform textual descriptions into architectural structures, or how different prompting strategies influence the resulting decompositions~\cite{WL33, 63}. This gap is particularly relevant given the growing reliance on documentation-driven decision-making in large-scale software evolution.

Taken together, the literature reveals both the potential and the limitations of current decomposition approaches. Traditional techniques offer well-established metrics and optimization strategies, but depend on artifacts that are often unavailable in early modernization stages. Conversely, LLM-based approaches provide new opportunities for abstraction and reasoning from textual inputs but lack systematic evaluation and methodological rigor. Understanding the behavioral patterns and trade-offs exhibited by LLMs in text-driven architecture decomposition is, therefore, a necessary step toward the design of more robust, hybrid, and reproducible decomposition pipelines. This motivation directly informs the methodology adopted in this study.

\subsection{Architecture-to-Architecture Transformations as a Research Problem}

Most existing work on software architecture modernization frames the transition from monoliths to microservices as a problem of \emph{reconstruction} or \emph{extraction}, typically grounded in source code analysis, runtime behavior, or expert-driven decision processes. In contrast, Architecture-to-Architecture transformations focus on the derivation of one architectural representation from another, often at a similar level of abstraction, without necessarily relying on implementation-level artifacts.

From a research perspective, Architecture-to-Architecture transformations introduce a distinct set of challenges. First, the absence of an objective ground truth makes correctness-based evaluation impractical, as multiple decompositions may be equally valid depending on architectural priorities and constraints. Second, architectural reasoning at this level is inherently interpretative, relying on abstractions, assumptions, and design heuristics rather than deterministic rules. Third, the transformation process itself may introduce implicit biases, reflecting the reasoning patterns or prior knowledge embedded in the transformation mechanism.

When driven by Large Language Models, these challenges are amplified. LLMs do not merely transform structure; they generate architectural artifacts through probabilistic reasoning over textual inputs. As a result, the generated architectures reflect not only the input descriptions, but also the model’s internal representations of architectural styles, modularity principles, and design trade-offs. Understanding these behaviors requires shifting the analytical focus from architectural optimality to architectural \emph{characterization}.

In this sense, text-driven Architecture-to-Architecture transformation can be viewed as a form of generative architectural reasoning, where the primary research question is not whether the resulting architecture is correct, but how different models and prompting strategies shape architectural structures, interaction patterns, and descriptive properties. This perspective motivates empirical studies that characterize LLM behavior through observable architectural signals, providing a foundation for more robust and guided architectural generation in future work.

\section{Research Methodology}
\label{sec:methodology}

To address the gaps identified in the literature—particularly the scarcity of empirical studies focusing on Architecture-to-Architecture transitions using Generative AI and the lack of reproducible evaluation mechanisms—this study adopts an exploratory and quantitative comparative analysis of architectural artifacts generated by Large Language Models (LLMs) from textual descriptions. We provide all artifacts used in this study, including textual requirements, prompting configurations, generated architectural outputs, and evaluation scripts, as supplementary material at\footnote{https://doi.org/10.6084/m9.figshare.31124914.v2}.

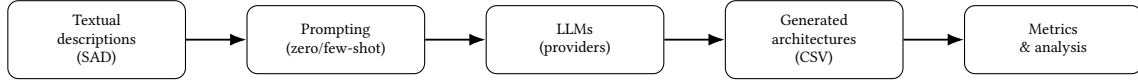
\begin{figure*}[t]
\centering
\begin{tikzpicture}[
  font=\scriptsize,
  node distance=7mm and 8mm,
  box/.style={draw, rounded corners, align=center, inner sep=5pt, minimum width=2.35cm, minimum height=0.9cm},
  arrow/.style={-Latex, line width=0.7pt}
]
\node[box] (input) {Textual\\descriptions\\(SAD)};
\node[box, right=of input] (prompt) {Prompting\\(zero/few-shot)};
\node[box, right=of prompt] (llm) {LLMs\\(providers)};
\node[box, right=of llm] (out) {Generated\\architectures\\(CSV)};
\node[box, right=of out] (eval) {Metrics\\\& analysis};

\draw[arrow] (input) -- (prompt);
\draw[arrow] (prompt) -- (llm);
\draw[arrow] (llm) -- (out);
\draw[arrow] (out) -- (eval);
\end{tikzpicture}
\caption{Overview of the study pipeline.}
\label{fig:pipeline_overview}
\end{figure*}

Figure~\ref{fig:pipeline_overview} provides an overview of the study pipeline, from textual inputs to artifact-based evaluation.

Framing the study as an Architecture-to-Architecture investigation implies a shift in evaluation focus. Rather than validating architectural correctness or optimality, the methodology emphasizes the characterization of structural and descriptive properties emerging from generative transformations between architectural representations at similar abstraction levels. This perspective aligns with the interpretative nature of early architectural reasoning, where multiple decompositions may coexist and architectural value depends on context, priorities, and trade-offs rather than absolute correctness.

Rather than relying on source code, runtime traces, or expert-driven evaluations, which dominate existing approaches to monolith decomposition, the proposed methodology investigates the architectural properties that emerge when LLMs are prompted with modular monolith requirements expressed exclusively in natural language. This design choice reflects early-stage modernization scenarios, where architectural decisions are often made based on partial documentation or high-level descriptions rather than complete implementation artifacts.

The study focuses on analyzing microservice architectures generated in a structured CSV format that explicitly captures service boundaries, declared responsibilities, and inter-service communication. By standardizing the output representation, the methodology enables systematic extraction of quantitative metrics and supports reproducibility across models and prompting strategies.

The decision to adopt a CSV-based representation is intentional and methodological rather than incidental. CSV files act as a lightweight architectural boundary object, offering a structured yet flexible representation that is sufficiently expressive to capture service boundaries, responsibilities, and interactions, while remaining model-agnostic and tool-independent. This choice facilitates automated parsing, metric extraction, and replication across different experimental settings, addressing reproducibility concerns frequently raised in recent GenAI-based software architecture studies.

We conducted a comparative experimental design to analyze the influence of two primary factors: (i) the choice of LLM provider and (ii) the prompting strategy. The evaluated systems (MediaStore and TeaStore) are treated as independent architectural scenarios, allowing observations to be replicated across distinct modular monolith contexts rather than serving as experimental factors themselves.

\textbf{Adaptation of the M2MDF Framework.} This study adopts and adapts the workflow proposed by the M2MDF. While the original framework prescribes a sequence of algorithmic steps for decomposition, we modify the standard pipeline to investigate the capabilities of LLMs. Traditional clustering or rule-based identification techniques are replaced with GenAI inference, while preserving the data preparation and evaluation principles established by the framework. 

\textbf{Data Acquisition and Pre-processing (Phases I \& II).} Unlike the standard Input Collection phase, which typically involves raw source code extraction, this research leverages a pre-validated dataset derived from prior literature~\cite{WL11}. We use the structured artifacts resulting from Static Analysis (SA) and Domain Analysis (DomA) performed in the baseline study by Fuchs et al.~\cite{WL11}. This approach provides a clean, abstract representation of monolithic systems, comprising components, responsibilities, and dependencies, thereby isolating the architectural reasoning task from the noise introduced by parsing low-level code. 

\textbf{Decomposition Strategy via Prompt Engineering (Phase III).} We conducted the Microservices Identification phase using LLMs under two experimental conditions. First, we applied a zero-shot prompting strategy, in which the model receives only the system description and decomposition objectives, allowing the assessment of its autonomous reasoning capabilities. Subsequently, we applied a few-shot prompting strategy, priming the model with example input-output decompositions to leverage in-context learning. This phase replaces unsupervised learning algorithms (e.g., K-means) commonly used in the M2MDF, shifting the focus to the models' semantic and structural reasoning.

\textbf{Evaluation and Metrics (Phase V).} Following the generation of candidate microservice architectures, the evaluation phase focuses on quantitative analysis of the produced artifacts. Due to the absence of architectural ground truth and implementation-level information, the generated solutions are not validated against a reference decomposition. Instead, evaluation is based on structural and textual properties extracted directly from the CSV representations.

Structural metrics characterize decomposition granularity and communication patterns, including the number of generated services, total and average inter-service communications, communication density, and the presence of isolated services. Complementarily, textual metrics assess the descriptive responsibility of services through measures such as average description length, vocabulary size, and lexical redundancy.

This artifact-based evaluation strategy enables systematic and reproducible comparison of different LLM providers and prompting strategies without requiring access to source code, runtime traces, or expert judgment. The Optimization and Deployment phases (Phases IV and VI of the original M2MDF) are considered outside the scope of this study, as the focus is exclusively on analyzing architectural properties emerging from text-driven microservice generation. 

It is important to emphasize that the selected metrics are deliberately limited to properties that can be reliably extracted from textual architectural artifacts. This constraint reflects both the exploratory nature of the study and the absence of implementation-level evidence. At the same time, it establishes a methodological baseline upon which richer evaluation layers can be incrementally integrated, including expert judgment, scenario-based trade-off analysis, or alignment with architectural quality attributes. In this sense, the current methodology is positioned as a first empirical step within a broader research agenda on hybrid and confidence-aware architectural evaluation using Generative AI.

\subsection{Rationale for Metric-Based Architectural Characterization}

Given the absence of implementation artifacts and reference decompositions, this study does not aim to assess architectural quality in normative terms. Instead, the adopted metrics serve as \emph{proxies} for architectural tendencies that emerge from LLM-generated artifacts. This distinction is essential to interpret the results appropriately.

Structural metrics such as the number of services, communication density, and isolated services provide indirect signals of how models reason about modularity, coupling, and service boundaries. Rather than indicating desirable or undesirable architectures, these measures reveal how different prompting strategies and model configurations influence decomposition behavior under comparable conditions.

Similarly, textual metrics related to responsibility descriptions are not intended to evaluate documentation quality per se. Instead, they capture the degree of elaboration, specificity, and lexical diversity produced by the models when articulating service responsibilities. These properties reflect how LLMs externalize architectural intent and may influence human interpretability during early design stages.

By grounding the evaluation on normalized, artifact-derived metrics, the study emphasizes reproducibility and comparability across experimental configurations. This approach aligns with recent calls in the literature for lightweight, artifact-centric evaluation strategies when studying Generative AI in architectural contexts where traditional validation methods are infeasible.

Importantly, this methodological choice positions the study as a foundational step toward more comprehensive evaluation pipelines. The characterized metrics can later be combined with expert judgment, scenario-based analyses, or implementation-level evidence, enabling progressive refinement of architectural assessment strategies in LLM-supported workflows.

\section{Results and Discussion}
\label{sec:results}

This section presents and discusses the quantitative results obtained from the generated microservice architectures. The analysis follows the experimental setup summarized in Table~\ref{tab:exp_config}, contrasting two LLM providers (OpenAI and DeepSeek) and two prompting strategies (zero-shot and few-shot) across two modular monolith projects. Given the absence of architectural ground truth, the discussion focuses on observable structural and textual properties directly derived from LLM outputs. In accordance with the evaluation strategy and metrics defined in Table~\ref{tab:metrics}, the goal is to characterize prompt- and model-driven patterns rather than architectural correctness.

\begin{table}[h]
\centering
\caption{Experimental Configuration}
\label{tab:exp_config}
\begin{tabular}{ll}
\hline
Aspect & Description \\
\hline
Architectural Style & Modular Monolith \\
Number of Projects & 2 \\
LLM Providers & OpenAI, DeepSeek \\
Prompting Strategies & Zero-shot, Few-shot \\
Runs per Configuration & 1 per project \\
Output Format & CSV \\
Evaluation Type & Quantitative (structural and textual) \\
\hline
\end{tabular}
\end{table}

\begin{table}[h]
\centering
\caption{Evaluation Metrics}
\label{tab:metrics}
\resizebox{\linewidth}{!}{%
\begin{tabular}{ll}
\hline
Metric & Description \\
\hline
Number of Services & Total number of generated microservices \\
Total Communications & Total declared inter-service communications \\
Average Communications & Mean number of communications per service \\
Communication Density & Ratio of declared communications to maximum possible \\
Isolated Services & Services with no declared communications \\
Avg. Responsibility Length & Mean number of words per responsibility description \\
Avg. Vocabulary Size & Mean number of unique words per responsibility \\
Redundancy Ratio & Proportion of services sharing frequent terms \\
\hline
\end{tabular}%
}
\end{table}

\subsection{Decomposition Granularity}

A clear distinction emerges between zero-shot and few-shot prompting strategies with respect to decomposition granularity. Consistent with the ``Number of Services'' metric described in Table~\ref{tab:metrics}, Table~\ref{tab:results_summary} shows that few-shot prompting consistently produces a higher number of microservices for both LLM providers, with reported averages reflecting aggregation across the two evaluated systems.

\begin{table*}[t]
\centering
\caption{Aggregated Results by Model and Prompting Strategy}
\label{tab:results_summary}
\begin{tabular}{llcccc}
\hline
Model & Setting & Services & Avg. Comms & Density & Avg. Resp. Length \\
\hline
DeepSeek & Zero-shot & 7.50 $\pm$ 0.71 & 3.07 $\pm$ 0.51 & 0.51 $\pm$ 0.08 & 9.27 $\pm$ 0.79 \\
DeepSeek & Few-shot  & 11.50 $\pm$ 0.71 & 1.32 $\pm$ 0.19 & 0.13 $\pm$ 0.02 & 11.97 $\pm$ 0.31 \\
OpenAI  & Zero-shot & 7.50 $\pm$ 0.71 & 3.21 $\pm$ 0.31 & 0.57 $\pm$ 0.00 & 14.63 $\pm$ 5.20 \\
OpenAI  & Few-shot  & 11.00 $\pm$ 0.00 & 1.18 $\pm$ 0.00 & 0.12 $\pm$ 0.00 & 19.06 $\pm$ 10.06 \\
\hline
\end{tabular}
\end{table*}

DeepSeek increases from an average of approximately 7.5 services in the zero-shot setting to over 11 services under few-shot prompting. A similar pattern is observed for OpenAI, which moves from around 7.5 to 11 services.

This behavior suggests that few-shot prompting encourages the models to adopt a more fine-grained interpretation of modular responsibilities. By providing an explicit example, the models appear more inclined to split functionality into smaller, more specialized services. While finer granularity is often desirable in some microservice contexts, excessive decomposition can also introduce additional coordination and communication overhead.

\subsection{Communication Structure and Density}

The communication-related metrics reveal an inverse relationship between service count and inter-service communication intensity. In particular, Table~\ref{tab:results_summary} indicates that zero-shot configurations exhibit higher average communication counts per service and higher communication density across both models, corresponding to the ``Average Communications'' and ``Communication Density'' metrics defined in Table~\ref{tab:metrics}.

For instance, zero-shot outputs reach densities above 0.5 in some configurations. In contrast, few-shot densities remain consistently closer to 0.12--0.15, indicating that architectures generated via zero-shot prompting are more tightly connected. In comparison, those generated via few-shot prompting are more loosely coupled.

Additionally, the few-shot configurations present a non-negligible number of isolated services, especially in the DeepSeek outputs. This observation aligns with the ``Isolated Services'' metric in Table~\ref{tab:metrics}, suggesting that finer-grained decompositions may introduce services with no declared interactions in the generated communication structure. These isolated services may represent utility-like components or reflect over-decomposition where responsibilities are split without corresponding interaction requirements.

Overall, these findings suggest that the prompting strategy directly affects not only the number of services generated, but also the distribution of architectural dependencies among them.

\subsection{Responsibility Descriptiveness}

Textual analysis of service responsibilities further highlights differences between prompting strategies. In line with the responsibility-focused metrics defined in Table~\ref{tab:metrics} (i.e., ``Avg. Responsibility Length'' and ``Avg. Vocabulary Size''), few-shot prompting produces longer responsibility descriptions with larger average vocabularies, particularly for OpenAI, where the average responsibility length nearly doubles compared to zero-shot outputs.

This increase in textual richness indicates that few-shot prompting encourages more detailed explanations of service responsibilities, potentially improving interpretability and documentation quality. However, longer descriptions do not necessarily imply better separation of concerns.

Despite the increased verbosity, the redundancy ratio remains high across all configurations. As captured by the ``Redundancy Ratio'' metric in Table~\ref{tab:metrics}, a large proportion of services share standard high-frequency terms in their responsibility descriptions, indicating substantial lexical overlap and suggesting that, even when services are decomposed more finely, the models tend to reuse similar domain terminology across multiple components.

\subsection{Cross-Model Comparison}

While both models show similar trends across prompting strategies, subtle differences emerge. OpenAI generally produces more verbose responsibility descriptions with higher vocabulary richness, particularly under few-shot prompting. In contrast, DeepSeek outputs show slightly higher variability in structural metrics, such as the presence of isolated services.

Nevertheless, the overall architectural patterns appear to be more strongly influenced by the prompting strategy than by the choice of LLM provider. This finding reinforces the importance of prompt design when using LLMs for architecture-related tasks.

\subsection{Cross-Metric Interpretation}

When considered jointly, the evaluated metrics defined in Table~\ref{tab:metrics} reveal a consistent trade-off between decomposition granularity, communication density, and responsibility descriptiveness. Few-shot prompting systematically favors finer-grained architectures with lower communication density and richer textual descriptions, whereas zero-shot prompting tends to produce coarser decompositions characterized by denser interaction structures.

The observed trade-offs suggest that LLMs implicitly encode architectural assumptions influenced by the provided prompting context. Few-shot examples appear to bias models towards modular expansion, whereas zero-shot configurations favor consolidating responsibilities. Importantly, neither behavior can be considered superior in isolation, as each reflects different architectural priorities that may be appropriate at various stages of system design. These observations should therefore be interpreted as tendencies in model behavior rather than as indicators of architectural quality or suitability for deployment.

\subsection{Limits of Metric-Driven Interpretation}

While the adopted metrics enable systematic comparison across models and prompting strategies, they also impose limitations on the interpretation of the results. Architectural properties such as service granularity or communication density are context-dependent and may reflect different design rationales rather than objective superiority or inferiority.

For instance, a higher number of services may indicate better separation of concerns in some contexts, but may also signal over-decomposition in others. Similarly, lower communication density can be interpreted as reduced coupling, but may also arise from underspecified interactions in early-stage designs. These ambiguities are inherent to Architecture-to-Architecture transformations, particularly when driven by generative models operating over textual descriptions.

Another limitation stems from the reliance on single-run configurations per experimental setup, which restricts the analysis to descriptive patterns rather than stability or variance estimation. Although the observed patterns are consistent across systems and models, repeated executions could reveal additional variability in decomposition outcomes. Such variability may itself be an important characteristic of LLM behavior, warranting explicit investigation in future studies.

These considerations reinforce that the reported metrics should be understood as descriptive signals rather than evaluative judgments. The primary contribution of this analysis lies in identifying systematic tendencies and trade-offs in LLM-generated architectures, which can inform both tool design and subsequent empirical investigations.

\section{Implications}
\label{sec:implications}

This study provides an initial empirical characterization of how Large Language Models behave when applied to text-driven Architecture-to-Architecture decomposition. By analyzing normalized structural and textual properties of generated microservice specifications, the results highlight systematic variations associated with prompting strategies and model configurations. Rather than offering prescriptive guidance, these findings help clarify observable tendencies in LLM behavior under constrained, early-stage architectural reasoning scenarios.

\textbf{Implications for Practice.} One implication concerns the scope of architectural evidence typically considered when evaluating LLM-generated designs. While the metrics adopted in this work capture relevant aspects of decomposition granularity, communication structure, and responsibility description, they represent only a partial view of architectural reasoning. Expanding the set of available signals to include static dependencies, runtime traces, or scenario-based trade-off analyses may enrich the interpretation of architectural structures and enable a more holistic assessment of generative outcomes.

\textbf{Implications for Research.} A second implication relates to the interplay between generative models and established architectural practices. The trade-offs observed between coarse-grained and fine-grained decompositions suggest that LLM outputs encode implicit architectural assumptions shaped by the prompting context, highlighting the opportunity to explore mechanisms that refine or steer these assumptions through prompt iteration, architectural constraints, or hybrid decision models, allowing for more controlled generation processes that align with domain-specific goals or architectural heuristics.

Finally, the observed sensitivity to prompt design reinforces that LLM-based architecture generation is best positioned as an exploratory and reflective activity, rather than a deterministic design mechanism. By surfacing alternative structural perspectives and decomposition patterns, LLMs can support architectural sensemaking and early design discussions. However, realizing this potential requires deeper empirical understanding of stability, variability, and alignment with human reasoning, which directly motivates the research agenda outlined next.

\textbf{Research Agenda for LLM-Supported Architectural Transformation.} 

The empirical patterns observed in this study suggest several directions for a broader research agenda on LLM-supported Architecture-to-Architecture transformation. First, future work should investigate the stability and variability of generated architectures under repeated executions, different temperature settings, and alternative prompt formulations. Understanding whether observed patterns persist or fluctuate across runs is essential for assessing the reliability of generative architectural reasoning.

Second, integrating complementary evidence sources represents a promising direction. Combining text-driven generation with static dependencies, runtime information, or architectural constraints may enable hybrid workflows that balance abstraction with technical grounding. Exploring how LLMs reconcile heterogeneous inputs remains an open research challenge.

Third, there is a need to examine alignment between LLM-generated architectures and human expert reasoning. Comparative studies involving architects could assess whether generated decompositions converge with expert intuition, diverge systematically, or reveal alternative but valid design perspectives.

Finally, future investigations should explore how characterized architectural signals can be operationalized in iterative design workflows. Rather than producing a single target architecture, LLMs could support architectural exploration by generating multiple alternatives, highlighting trade-offs, or supporting what-if analyses during early modernization phases.

Taken together, these directions define a progressive research agenda in which LLMs are not positioned as autonomous architects, but as generative partners whose behaviors, limitations, and strengths must be empirically characterized. This study contributes an initial empirical baseline for that agenda, clarifying how architectural structures emerge from text-driven prompts and establishing foundations for more controlled, hybrid, and human-centered architectural transformation workflows.

\section{Threats to Validity}
\label{sec:threats}

This study is subject to several threats to validity, which are discussed following the classification proposed by Wohlin et al.~\cite{wohlin2012experimentation}.

\textbf{Internal Validity}. Concerns factors that may influence the observed results independently of the studied variables. The main threat arises from the inherent stochasticity and LLMs' sensitivity to prompt formulation. Although prompt templates were applied consistently across models and systems, single-run executions limit our ability to assess generation variance. Replicating configurations across multiple seeds or evaluating output stability across repeated runs could help characterize LLM behavior variability more systematically. Importantly, variability should not be interpreted solely as a threat, but also as an object of study in its own right, as it may reveal systematic behavioral properties of generative architectural reasoning.

\textbf{Construct Validity}. Refers to the extent to which the employed metrics capture the intended architectural properties. Our evaluation focuses on structural and textual features extracted from CSV artifacts. While these metrics are reproducible and well-defined, they do not directly assess quality attributes such as maintainability or scalability. Incorporating scenario-based assessments, domain-specific quality indicators, or richer architectural representations may enable deeper insights in future investigations.

\textbf{External Validity}. Addresses the generalizability of the findings. The study considers two modular monoliths and two LLM providers, which limits its breadth. The exclusive use of textual descriptions reflects early-stage modernization contexts but may not generalize to scenarios with access to source code or runtime data. Exploring additional systems, domains, or architectural paradigms could strengthen external validity and support broader generalization.

\textbf{Conclusion Validity}. The limited number of systems and single-run configurations restricts the statistical power of the reported findings. As the study is exploratory, we refrain from formal hypothesis testing. However, complementing the current metrics with statistical analyses or expert-coded evaluations could enhance the interpretability and robustness of conclusions in future work.

Despite these threats, the study adopts a transparent and reproducible design, and the consistency of the observed patterns across LLMs and prompting strategies suggests that the findings capture systematic behavioral traits rather than isolated phenomena.

\section{Final Remarks}
\label{sec:finalremarks}

This study examined the use of LLMs to derive microservice architectures from textual descriptions of modular monoliths, addressing a gap in the literature related to Architecture-to-Architecture transformations supported by Generative AI. Unlike approaches that rely on source code, runtime traces, or expert-driven validation, the investigation focused exclusively on architectural artifacts generated from natural language prompts, reflecting early-stage modernization scenarios.

The results indicate that prompting strategies systematically influence the structural and descriptive properties of the generated architectures. Few-shot prompting tended to produce finer-grained decompositions with lower communication density and more detailed responsibility descriptions, whereas zero-shot prompting favored coarser structures with denser interaction patterns. These tendencies were observed consistently across two LLM providers, underscoring the central role of prompt design in text-driven architectural generation. Importantly, these results should not be interpreted as evidence of architectural quality or suitability for deployment. Instead, they provide descriptive insights into how LLMs operationalize architectural concepts when reasoning over textual representations under controlled conditions.

Rather than claiming architectural correctness or optimality, this work provides empirical observations on how LLMs behave when applied to decomposition tasks under constrained input conditions. By adopting normalized, artifact-based metrics derived from standardized representations, the study provides a lightweight, reproducible lens for characterizing generative outputs in the absence of implementation-level artifacts.

Beyond these findings, the proposed methodology provides a foundation for broader investigations into how LLMs can support architectural reasoning. The experimental setup, artifact structure, and evaluation pipeline can be extended to include more diverse inputs, incorporate domain-specific constraints, or integrate expert feedback. As interest in hybrid and semi-automated software architecture workflows grows, the ability to systematically analyze and compare generative behaviors becomes essential. By positioning LLMs as exploratory partners rather than decision-makers, future work can build on the foundations established here to develop more adaptable, explainable, and human-centered architectural transformation processes.

\bibliographystyle{ACM-Reference-Format}
\bibliography{references.bib}

\end{document}